\documentclass[aps, prl, reprint, superscriptaddress]{revtex4-1}
\usepackage{graphicx,amssymb,amsfonts,amsmath,subfigure, upgreek}
\usepackage{natbib}
\usepackage{chemformula}
\usepackage{color,soul}
\usepackage{xspace}
\usepackage{xcolor}
\emergencystretch=\maxdimen
\begin{document}
\title{Evidence of Magnon-Mediated Orbital Magnetism in a Quasi-2D Topological Magnon Insulator}

\author{Laith Alahmed}
\affiliation{Department of Electrical and Computer Engineering, Auburn University, Auburn, AL 36849, USA}

\author{Xiaoqian Zhang}
\affiliation{Shenzhen Institute for Quantum Science and Engineering, Southern University of Science and Technology, Shenzhen, 518055, China}

\author{Jiajia Wen}
\affiliation{Stanford Institute for Materials and Energy Sciences, SLAC National Accelerator Laboratory, Menlo Park, CA, 94025, USA}

\author{Yuzan Xiong}
\affiliation{Department of Physics, Oakland University, Rochester, MI 48309 USA} 

\author{Yi Li}
\affiliation{Materials Science Division, Argonne National Laboratory, Lemont, IL 60439}

\author{Li-chuan Zhang}
\affiliation{Peter Gr\"unberg Institut and Institute for Advanced Simulation,
Forschungszentrum J\"ulich and JARA, 52425 J\"ulich, Germany}

\author{Fabian Lux}
\affiliation{Institute of Physics, Johannes Gutenberg University Mainz, 55099 Mainz, Germany}

\author{Frank Freimuth} \affiliation{Peter Gr\"unberg Institut and Institute for Advanced Simulation,
Forschungszentrum J\"ulich and JARA, 52425 J\"ulich, Germany}
\affiliation{Institute of Physics, Johannes Gutenberg University Mainz, 55099 Mainz, Germany}

\author{Muntasir Mahdi}
\affiliation{Department of Electrical and Computer Engineering, Auburn University, Auburn, AL 36849, USA}

\author{Yuriy Mokrousov}
\affiliation{Peter Gr\"unberg Institut and Institute for Advanced Simulation,
Forschungszentrum J\"ulich and JARA, 52425 J\"ulich, Germany}
\affiliation{Institute of Physics, Johannes Gutenberg University Mainz, 55099 Mainz, Germany}

\author{Valentine Novosad}
\affiliation{Materials Science Division, Argonne National Laboratory, Lemont, IL 60439}

\author{Wai-Kwong Kwok}
\affiliation{Materials Science Division, Argonne National Laboratory, Lemont, IL 60439}

\author{Dapeng Yu}
\affiliation{Shenzhen Institute for Quantum Science and Engineering, Southern University of Science and Technology, Shenzhen, 518055, China}

\author{Wei Zhang}
\affiliation{Department of Physics, Oakland University, Rochester, MI 48309 USA}

\author{Young S. Lee}
\affiliation{Stanford Institute for Materials and Energy Sciences, SLAC National Accelerator Laboratory, Menlo Park, CA, 94025, USA}
\affiliation{Department of Applied Physics, Stanford University, Stanford, CA, 94305, USA}

\author{Peng Li}
\affiliation{Department of Electrical and Computer Engineering, Auburn University, Auburn, AL 36849, USA}

\begin{abstract}
\noindent\textbf{ABSTRACT:} We explore spin dynamics in Cu(1,3-bdc), a quasi-2D topological magnon insulator. The results show that the thermal evolution of Landé $g$-factor ($g$) is anisotropic: $g_\textrm{in-plane}$ reduces while $g_\textrm{out-plane}$ increases with increasing temperature $T$. Moreover, the anisotropy of the $g$-factor ($\Delta g$) and the anisotropy of saturation magnetization ($\Delta M_\textrm{s}$) are correlated below 4 K, but they diverge above 4 K. We show that the electronic orbital moment contributes to the $g$ anisotropy at lower $T$, while the topological orbital moment induced by thermally excited spin chirality dictates the $g$ anisotropy at higher $T$. Our work suggests an interplay among topology, spin chirality, and orbital magnetism in Cu(1,3-bdc).

\end{abstract}

\maketitle
The discovery of topologically protected states in some systems with fermionic particles (e.g. electrons and holes) led to extensive research on topological insulators unraveling their unique properties\cite{RevModPhys.82.3045}. Such topology-protected states can also exist within the band gap of systems with bosonic quasi-particles, such as photons~\cite{haldane2008possible,raghu2008analogs}, phonons~\cite{LiHuang18} and magnons~\cite{PhysRevB.87.144101,Malki_2020,Pereiro2014,Malki2019PRB,PhysRevB.100.144401,mook2020chiral}, which can mediate the transport of spin and orbital angular momentum~\cite{kovalev2016spin,cheng2016spin,zhang2020imprinting,zhang2010topological}. While the interplay between the topology of electronic bands and spin transport properties has been intensively studied~\cite{Li2018,Niyazov2020,PhysRevResearch.3.023219}, the relationship between magnonic topology and intrinsic magnetic properties remains largely unexplored~\cite{Pereiro2014}.

Non-trivial magnonic band topology was predicted in magnonic crystals such as Lu$_2$V$_2$O$_7$ and Cu[1,3- benzenedicarboxylate(bdc)] \cite{PhysRevB.87.144101}. Cu(1,3-bdc) is a metal-organic hybrid material where Cu$^{2+}$ ions are arranged in a geometrically perfect Kagome lattice structure. The organic (1,3-bdc) molecules separate the individual Kagome planes, leading to weak interlayer interaction. It is thus identified as the first quasi-two-dimensional (2D) topological magnon insulator where strong exchange coupling is confined within individual layers~\cite{Nytko2008}. Recent neutron scattering experiments identified flat bands originating from the unique geometry of the Kagome lattice, which can be described by a Heisenberg Hamiltonian with Dzyaloshinskii-Moriya interaction~\cite{PhysRevLett.115.147201,PhysRevB.93.214403}. Thus, these exotic properties make Cu(1,3-bdc) an ideal material platform for exploring the interplay between magnonic topology and intrinsic magnetic properties including magnetization dynamics~\cite{Malz2019}.

A recent theory proposed that chiral magnetism and topological magnonic excitations can be correlated with electronic orbital magnetism~\cite{HuachenPRB,PhysRevLett.125.117209}. In particular, it is suggested that orbital magnetization can play a significant role in the dynamics of collinear antiferromagnets with weak spin-orbit coupling~\cite{HuachenPRB}. However, experimental evidence concerning the role of magnon-mediated electronic orbital moment in magnetization dynamics is still lacking. In this regard, it is crucial to carry out experiments to uncover the underlying physics on this topic.

\begin{figure}[!t]
    \centering
    \includegraphics[width=0.5\textwidth]{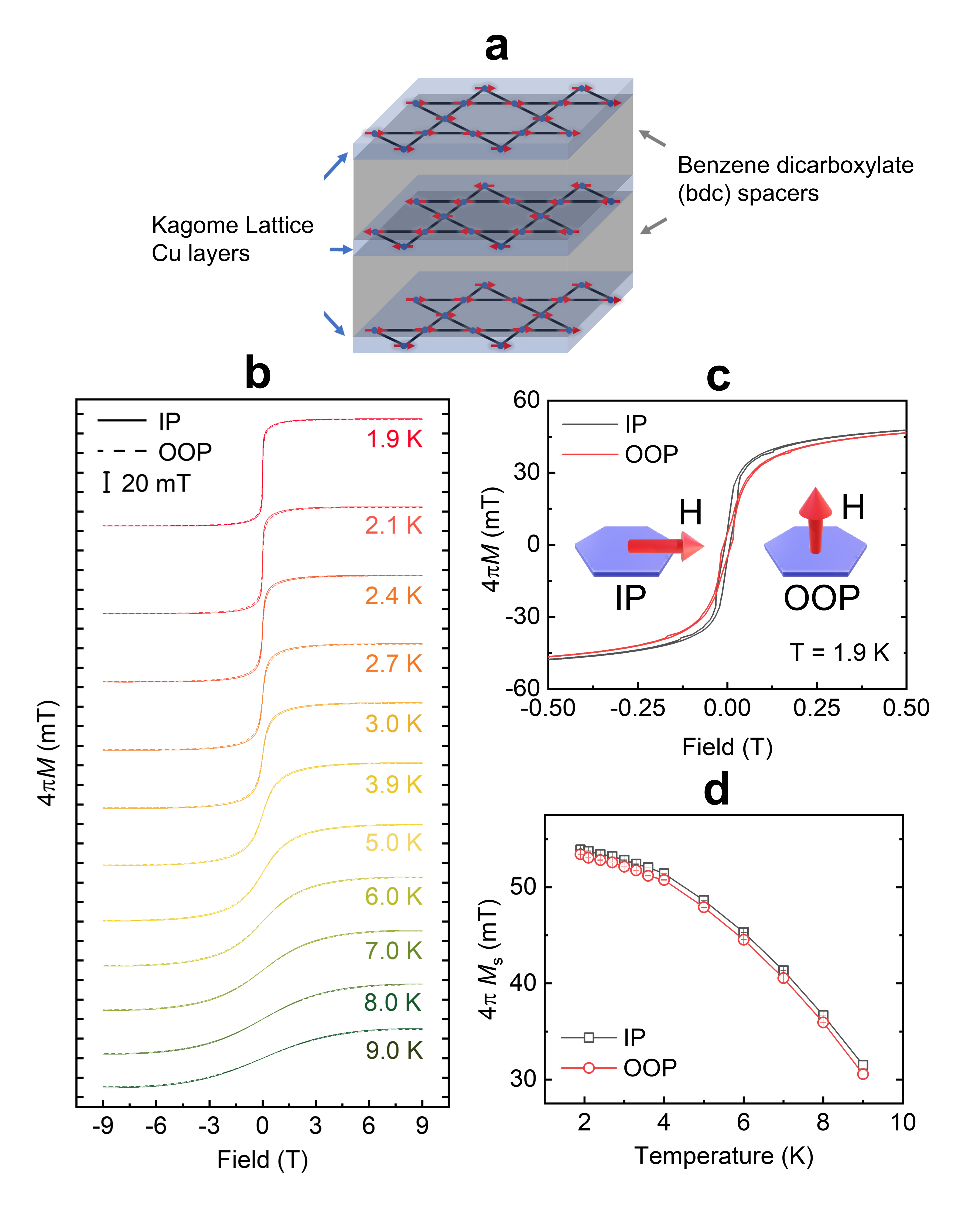}
    \caption{Quasi-static magnetic properties of Cu(1,3-bdc). \textbf{a.} Schematic of Cu(1,3-bdc) structure. Kagome lattice Cu layers are spaced by Benzene dicarboxylate (bdc) organic layers. Red arrows represent spins within the Kagome lattice. \textbf{b.} Magnetization vs. externally applied field plots at different temperatures. The magnetic field was applied along the Kagome plane (IP) of the Cu(1,3-bdc) sample (solid lines), and perpendicular (OOP) to the Kagome plane (dashed lines). \textbf{c.} Zoomed-in IP and OOP hysteresis loops at T = 1.9 K, showing IP anisotropy. The left and right insets show the IP and OOP field directions, respectively. \textbf{d.} Temperature dependence of $M_\textrm{s,IP}$ (gray) and \textit{M}$_\textrm{s,OOP}$(red) extracted from the VSM measurements.}
    \label{fig1}
\end{figure}

Here, we use Cu(1,3-bdc) as a model material to study the correlation of Landé $g$-factor and orbital angular momentum through Vibrating Sample Magnetometry (VSM), Superconducting Quantum Interference Device (SQUID), and broadband Ferromagnetic Resonance (FMR) spectroscopy. The experimental results show an anisotropic gyromagnetic ratio and a corresponding $g$-factor tensor. This anisotropy is found to be correlated with the difference in In-Plane (IP) and Out-Of-Plane (OOP) saturation magnetization at lower temperatures (T \textless 4 K); it indicates a contribution from electronic orbital moment. Surprisingly, such correlation breaks down when T \textgreater 4 K. Further analysis reveals that magnon-mediated electronic topological orbital moment has contributed to the magnetic moment at higher temperatures. We have built a foundation for elucidating the intriguing interplay of topology, spin excitations, and electronic orbital magnetism. This work also highlights the unique properties of quasi-2D topological magnon insulators for building novel spintronic devices.

It was previously shown that, in the ground state of Cu(1,3-bdc), spins within each Kagome plane are ferromagnetically ordered, while spins in neighboring planes are antiferromagnetically ordered (Fig. 1a). The interlayer antiferromagnetic coupling was found to be $\sim$ 0.3\% of the intralayer nearest-neighbor coupling. Thus, the magnetization across different Kagome planes can be easily aligned by a weak magnetic field ($\sim$ tens of mT)\cite{PhysRevB.93.214403}.

The quasi-static magnetization data of Cu(1,3-bdc) are presented in Fig. \ref{fig1}. Here, several notes should be made: First, the magnetization as a function of IP and OOP fields at temperatures ranging from 1.9 K to 9 K was measured (Fig. 1a). Consistent with the previous measurement\cite{PhysRevB.93.214403}, Cu(1,3-bdc) favors an slight in-plane alignment. Second, Cu(1,3-bdc) is a quasi-2D material. It is estimated that the inter-Kagome-plane interaction is about 0.3$\%$ of the intra-Kagome-plane interaction. The three-dimensional Néel transition at $\sim$1.8 K corresponds to the antiferromagnetic ordering of the spins on different Kagome planes, which is driven by the weak inter-plane interaction. The dominant intra-Kagome-plane interaction, which is ferromagnetic, is responsible for the short-range ferromagnetic ordering within each plane that persists well above the transition temperature of 1.8 K\cite{PhysRevB.93.214403}. Thus, we summarize the magnetic transitions in Cu(1,3-bdc) as follows: when $T<$1.8 K, Cu(1,3-bdc) exhibits antiferromagnetic ordering of the spins on different Kagome planes. When 1.8 K$<T<$9 K, short-range ferromagnetic ordering exists within each plane. when T$>$9 K, it transitions into a more paramagnetic state. Third, the magnetization curves were fitted with a hyperbolic tangent function to extract the saturation magnetization $M_\textrm{s}$ \cite{Li2021} (refer to Supplementary Note V for details). The temperature dependence of the IP and OOP saturation magnetization $M_\textrm{s,IP}$, $M_\textrm{s,OOP}$ is plotted in Fig. \ref{fig1}c. Intriguingly, $M_\textrm{s,IP}$ is higher than that of $M_\textrm{s,OOP}$ across all tested temperatures (1.9 K to 9 K). For example, $M_\textrm{s,IP}$ is $\sim$0.8$\%$ larger than $M_\textrm{s,OOP}$ at 1.9 K, which has been confirmed by both the VSM and SQUID measurements.  This effect is also observed in the previous magnetization measurement\cite{PhysRevB.93.214403}. With the hypothesis that the difference in $M_\textrm{s,IP}$ and $M_\textrm{s,OOP}$ resulted from orbital magnetic moment\cite{HuachenPRB,PhysRevLett.125.117209}, we carry out magnetization dynamics measurements to verify this understanding.

\begin{figure}[!t]
    \centering
    \includegraphics[width=0.5\textwidth]{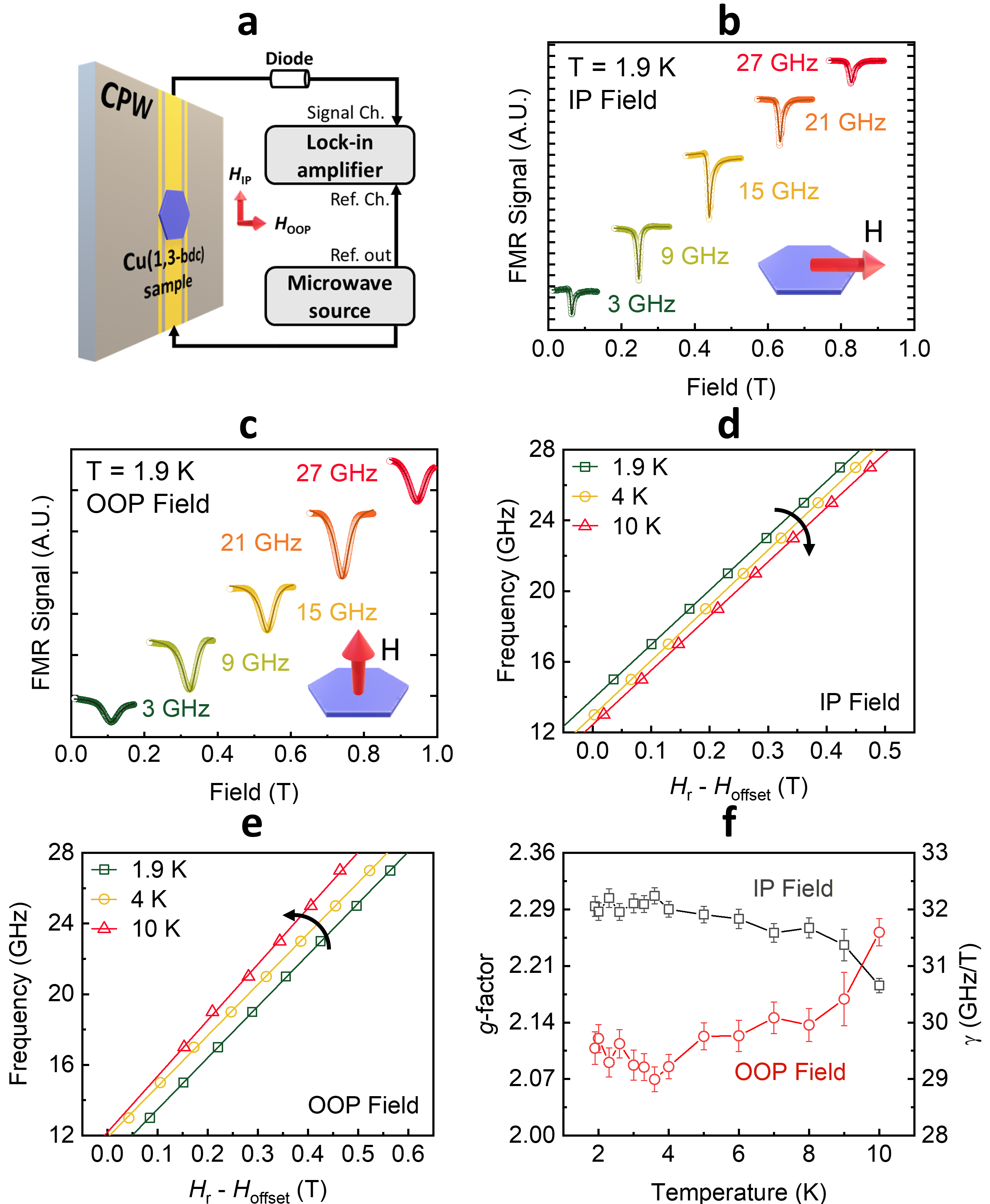}
    \caption{Broadband FMR spectroscopy and analysis. \textbf{a.} Schematic of the experimental setup. \textbf{b.} and \textbf{c.} FMR profiles measured at different microwave frequencies at T = 1.9 K for IP and OOP fields, respectively. Both the datapoints and fitted curves are presented. The insets in \textbf{b} and \textbf{c} show the IP and OOP field directions, respectively. \textbf{d.} and \textbf{e.} Resonance frequency \textit{f} vs. resonance field \textit{H}$_\textrm{r}$-\textit{H}$_\textrm{offset}$ at 1.9 K, 4 K, and 10 K for external fields applied IP and OOP, respectively. Here, \textit{H}$_\textrm{offset}$ = 0.4 T. The datapoints are fitted to Kittel Equations 1 and 2, respectively. The arrows denote the behavior of increasing temperature. \textbf{f.} Temperature dependence of the Landé \textit{g}-factor (left axis) and gyromagnetic ratio $\gamma$ (right axis) for external fields along the IP (gray) and OOP (red) directions.}
    \label{fig2}
\end{figure}
   
Fig. \ref{fig2} presents the FMR data. A schematic of the FMR measurement is shown in Fig. \ref{fig2}a. A microwave diode and lock-in amplifier were used for signal detection. A Vector Network Analyzer (VNA) was also used to perform the FMR measurements and generated consistent data (See Supplementary Note 7 for details). Figs. \ref{fig2}b,c show microwave power absorption vs. external field at microwave frequencies ranging from 3 GHz to 27 GHz at 1.9 K along the IP and OOP directions, respectively. The FMR profiles were measured at temperatures up to 10 K. At each microwave frequency \(f\), the FMR profiles were fitted to Lorentzian + anti-Lorentzian functions\cite{Oates2002,LowYIG} to extract the resonance field $H_\textrm{r}$, and the Full Width at Half Maximum (FWHM) linewidth $\Delta H_\textrm{FWHM}$ (See Supplementary Note 8). The resonance frequencies are plotted against the fitted IP and OOP resonance fields as shown in Figs. \ref{fig2}d,e. We adopt $H_\textrm{r}$-$H_\textrm{offset}$ as the horizontal axes, where $H_\textrm{offset}$ = 0.4 T. These plots reveal an linear relationship between the resonance frequency and the resonance magnetic field (refer to Supplementary Note VI for details). The data points can be fitted by the Kittel equations\cite{LowYIG}:

\begin{equation}
f = \gamma'_\textrm{IP} \sqrt{(H_r +4\pi M_\textrm{eff}) H_r} \label{Kittel_IP}
\end{equation}
\begin{equation}
f = \gamma'_\textrm{OOP} (H_r -4\pi H_\textrm{eff}) \label{Kittel_OOP}
\end{equation}

Equations (\ref{Kittel_IP}) and (\ref{Kittel_OOP}) are for the IP and OOP cases, respectively. Here, $f$ is the microwave frequency, $\gamma'$ = $\frac{\gamma}{2\pi}$ is the reduced gyromagnetic ratio, and $H_\textrm{eff}$ is the effective field representing the contribution from magnetocrystalline anisotropy. The $g$-factor is calculated using $g=|\gamma|\frac{\hbar}{\mu_\textrm{B}}$, where $\mu_\textrm{B}$ is the Bohr magneton and $\hbar$ is the reduced Planck constant.

Fig. \ref{fig2}f plots the temperature dependence of the $g$-factor (left Y-axis) and $\gamma'$ (right Y-axis) along the IP and OOP fields, respectively. Several observations can be made from Fig. \ref{fig2}f: First, in both IP and OOP cases, the $g$-factor deviates from $g$ = 2 that is for a free electron without orbital momentum. This deviation indicates an orbital contribution to the magnetic moment in Cu(1,3-bdc)\cite{PhysRevB.100.134437,FeGeTe}. Second, $g_\textrm{IP}$ is significantly larger than $g_\textrm{OOP}$ at lower temperatures (i.e., \textless 9 K). Third, the thermal evolution of $g_\textrm{IP}$ and $g_\textrm{OOP}$ shows opposite trends: $g_\textrm{IP}$ reduces with increasing temperature, while $g_\textrm{OOP}$ increases with increasing temperature.

Both the quasi-static and dynamic magnetization measurements suggest the footprint of electronic orbital moment. Thus, the difference of the $g$-factor $\Delta g$ = $g_{\textrm{IP}}$-$g_{\textrm{OOP}}$ is plotted together with the difference of the saturation magnetization $\Delta M_\textrm{s}$ = $M_{\textrm{s,IP}}$-$M_{\textrm{s,OOP}}$ in Fig. \ref{Ham}a. Fig. \ref{Ham}a shows two contrasting behaviors, a low-temperature regime where the behavior of $\Delta g$ is similar to $\Delta M_\textrm{s}$ with no significant deviation at T \textless 4 K, and a high-temperature regime where $\Delta g$ deviates greatly from $\Delta M_\textrm{s}$ at T \textgreater 4 K.

It is thus insightful to interpret the temperature-dependent anisotropy of the $g$-factor in terms of the dynamics of orbital magnetism. At low temperatures, the $g$-factor anisotropy can be attributed to the anisotropy of the orbital magnetization in Cu(1,3-bdc) along the IP and OOP directions, with the IP case having a greater orbital contribution. The deviation of $\Delta g$ from $\Delta M_\textrm{s}$ at higher temperatures comes as a surprise, because they are expected to be closely correlated at all temperatures. Thus, this points to a different mechanism to the orbital correction of the $g$-factor; it cannot be explained by the orbital motion of electrons around atomic cores.

A recent theoretical work uncovered contributions to electronic orbital magnetism originating from spin chirality generated from spin disorder\cite{zhang2020imprinting, wimmer2019chirality}. This new development in the area of thermally-driven spin fluctuations points out a route to explain the observed behavior in Cu(1,3-bdc). According to the theory, the non-vanishing net spin chirality can arise even in a collinear fluctuating spin system in its ground state. This can be directly translated into a topological electronic orbital motion, the strength of which is given by the topological orbital susceptibility relating the degree of the chirality to the magnitude of the orbital magnetization~\cite{grytsiuk2020topological}. The magnitude of the topological orbital susceptibility can be sizable in materials with weak spin-orbit coupling, i.e. Cu(1,3-bdc), as shown in conventional microscopic calculations. Furthermore, in ferromagnetic Kagome systems with non-trivial topological magnonic bands, the spin chirality mediated by thermal fluctuations can imprint sizeable electronic orbital magnetization; their sign and magnitude are controlled by the parameters of the studied system~\cite{zhang2020imprinting}.

Following the approach in Ref.~\cite{zhang2020imprinting}, we use the extracted parameters which describe the spin-exchange interactions in Cu(1,3-bdc)~\cite{PhysRevLett.115.147201} to compute the magnonic properties. The calculation yielded thermally induced net chirality and corresponding orbital magnetization in the system. Specifically, we use the effective spin Hamiltonian of Cu(1,3-bdc) in Eq.(\ref{Ham}):

\begin{equation}
\begin{split}
    H=&-\frac{1}{2}\sum_{ij}J_{ij}  \mathbf{S}_i \cdot \mathbf{S}_j -\frac{1}{2}\sum_{ij}\mathbf{D}_{ij} \cdot (\mathbf{S}_i\times \mathbf{S}_j)\\  &-\mathbf{B}\cdot \kappa^\mathrm{TO}\sum_{ijk} \hat{\mathbf e}_{ijk} 
    [
    \hat{\mathbf{S}}_i \cdot (
    \hat{\mathbf{S}}_j \times
    \hat{\mathbf{S}}_k)] 
    -{{\mu_\mathrm B}}g_e \mathbf{B}\cdot\sum_i \mathbf{S}_i  \, ,
\end{split}
\label{Ham}
\end{equation} 

where $J_{\textrm{ij}}$  mediates the Heisenberg exchange between spins $\mathbf{S}_{\textrm{i}}$ and $\mathbf{S}_{\textrm{j}}$ on sites $i$ and $j$ in the first term. The second term is the antisymmetric Dzyaloshinskii-Moriya Interaction (DMI) quantified by vectors $\mathbf{D}_{\textrm{ij}}$. The fourth term couples the spins to an external magnetic field $\mathbf B$. The third term is the ring-exchange interaction term, which explicitly describes the Zeeman interaction of the topological orbital moment (TOM, $\mathbf L^{\textrm{TOM}}$) with the external magnetic field $\mathbf B$~\cite{zhang2020imprinting}. TOM marks a special type of electronic orbital moment. This term is given by the product of the scalar spin chirality (SSC) and the topological orbital susceptibility $\kappa^\mathrm{TO}$~\cite{wimmer2019chirality,hanke2017prototypical,grytsiuk2020topological,hanke2016role}. Owing to the symmetry of the planar Kagome lattice, both the TOM and DMI vectors are perpendicular to the Kagome plane (OOP), in the same direction as the applied external magnetic field $\mathbf B$. In Eq.(\ref{Ham}), the  spin operator $\mathbf S_\textrm{i}$ at site $i$ is set as $\frac{1}{2}$, $g_\textrm{e}$ is set as 2, the nearest-neighbor exchange interaction $J$ is chosen as 0.6~meV and the DMI vector is set as $D_{\textrm{ij}}=(0,0, 0.09)$~meV. The topological orbital susceptibility $\kappa^{\textrm{TO}}$ is chosen as a typical value of $-0.2\mu_\textrm{B}$, as suggested by the previous studies~\cite{wimmer2019chirality,hanke2017prototypical,grytsiuk2020topological,hanke2016role}. 

\begin{figure}[t!]
\centering
\includegraphics[width=0.5\textwidth]{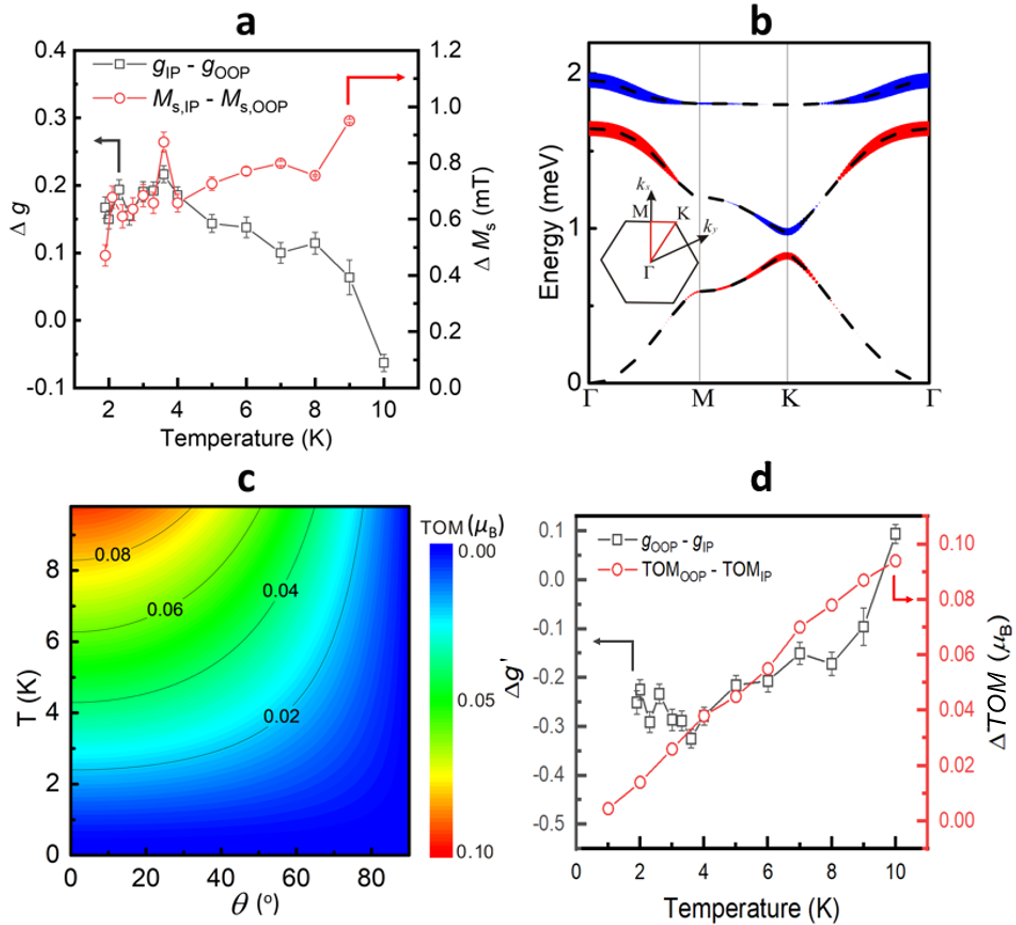}
\caption{Electronic Topological Orbital Moment in Cu(1,3-bdc). \textbf{a.} Temperature dependence of $\Delta g$ (plotted on the left Y-axis) and $\Delta$\textit{M}$_\textrm{s}$ (plotted on the right Y-axis). \textbf{b.} Flat band analysis for the magnonic bands of Cu(1,3-bdc). Red and blue colors represent the positive and negative signs of the local topological orbital moment (TOM) $\mathbf L^{\textrm{TOM}}$, respectively. The line thickness denotes the corresponding magnitude.  The insert represents the first Brillouin Zone. The red lines connect the high symmetry points which are selected in the dispersion. \textbf{c.} TOM magnitude vs. temperature and polar angle $\theta$. $\theta$=0$^o$ (90$^o$) means that magnetization is along the OOP (IP) direction. The colors represent the magnitude of the integrated TOM in $\mu_\textrm{B}$. Magnetic field $B$ is assumed to be zero. \textbf{d.} $\Delta g'$ (left Y-axis) and $\Delta$TOM (right Y-axis) as a function of temperature.}
\label{tom}
\end{figure}

In Fig. \ref{tom}b, we plot the magnonic band-resolved contributions to the orbital moment in Cu(1,3-bdc) for $B$ = 0. One can see a strong correlation between the orbital moment and the topological band inversion of magnonic bands at low energies and between the first and second modes. Then, we integrate the contribution of the band-resolved orbital moment to the overall orbital moment of Cu(1,3-bdc) at finite temperatures. Fig. \ref{tom}c presents the calculated TOM as a function of temperature and polar angle $\theta$ that indicates the magnetization direction. From Fig. \ref{tom}c, two observations are made: First, the analysis shows that the symmetry of the system allows orbital magnetization originating from spin excitations in the OOP case but prohibits it in the IP case. Second, the value of the thermally-induced orbital moment increases monotonously with temperature and becomes sizable when  T \textgreater 4 K along the OOP direction. In Fig. \ref{tom}d, the differences of the $g$-factor ($\Delta g' = g_\textrm{OOP}-g_\textrm{IP}$) and ($\Delta TOM = TOM_\textrm{OOP}-TOM_\textrm{IP}$) are plotted as a function of temperature. One can see that the $\Delta g'$ and $\Delta TOM$ deviate below 4 K but become strongly correlated above 4 K. Thus, our theory accurately predicts the general trend and the change in the sign of $\Delta g$ observed experimentally, as shown in Fig. \ref{tom}d. These facts suggest that the spin excitations can renormalize the fundamental quantum mechanical constant conventionally associated with the atomic orbital magnetism. To this end, our findings have provided experimental evidence of a new mechanism for orbital dynamics in topological magnonic systems.

\begin{figure}[!t]
    \centering
    \includegraphics[width=0.5\textwidth]{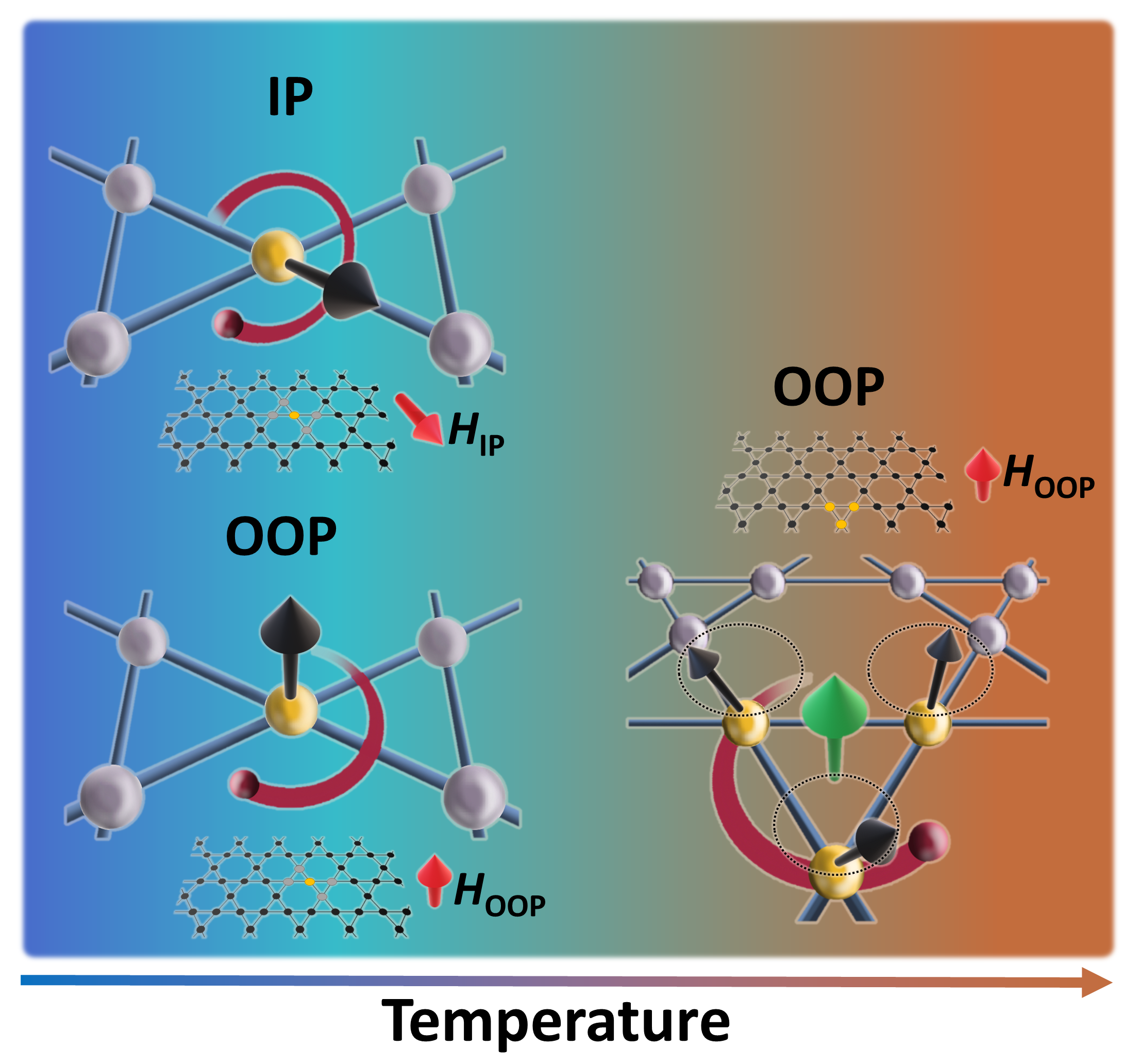}
    \caption{Orbital Magnetic Moment. The left two illustrations show the electronic orbital moment occurring at low temperatures. The black arrows represent the overall spin and orbital moment for a field applied IP (top left) and OOP (bottom left). The right illustration shows the magnon-mediated topological orbital moment (green arrow) arising from thermally-driven scalar spin chirality.}
    \label{fig4}
\end{figure}

Figure \ref{fig4} summarizes all the orbital contributions to the magnetic moment. At low temperatures, the orbital contributions come from the motions of electrons orbiting their atomic cores for both IP and OOP cases. The black arrows in the left two illustrations represent the summation of the spin and the orbital angular momentum of the orbiting electron (red) around the nucleus (yellow). When temperature increases, the thermal fluctuations give rise to spin chirality allowed by the symmetry of the Kagome lattice in the OOP case (as shown in the right illustration). In this scenario, an electron (red) hops between a precessing noncolinear spin-triplet (yellow), generating TOM (green arrow). As temperature increases, magnon scattering, such as magnon-magnon scattering and magnon-phonon scattering, causes IP and OOP orbital moment to decay. This is why the $g$-factor along the IP direction reduces at higher temperatures, as shown in Fig. \ref{fig2}f. In this regard, the increase of the $g$-factor along OOP with increasing temperature highlights the significance of TOM in the Cu(1,3-bdc) system.

It is helpful to identify the limiting case of the TOM effect, namely, the saturation temperature of TOM. The linear spin-wave theory we have used cannot predict the saturation temperature of TOM. Thus, we have explored the $g$-factor above 10 K using the FMR technique. Our results show that the FMR signals become significantly weaker and nosier above 10 K. We have estimated the $g_\textrm{OOP}$ and $g_\textrm{IP}$ parameters, which decay and approach $g$ = 2 beyond 12 K. This indicates that TOM may saturate around 12 K.

In summary, anisotropies of Landé $g$-factor and saturation magnetization are observed in Cu(1,3-bdc). When $T$ \textless 4 K, the differences of $g$-factor ($\Delta g$) and saturation magnetization ($\Delta M_\textrm{s}$) are correlated, which indicates the contribution of orbital moment of electrons to the magnetic moment. The deviation of $\Delta g$ and $\Delta M_\textrm{s}$ at T \textgreater 4 K can be explained by the spin chirality mediated by thermal fluctuations inducing sizeable electronic orbital magnetization. Our work has highlighted Cu(1,3-bdc) as an important material platform to understand the interplay of topology, spin excitations and orbital magnetism, thereby presenting potential direction for establishing new material platforms for building novel spintronic devices. Moreover, our study has pointed a new way to probe the orbital moment in quantum magnets via the $g$-factor measurements. Future work that studies the chiral and topological orbital magnetism of domain walls and skyrmions\cite{lux2018engineering}, as well as skyrmion-topological magnon interactions in Cu(1,3-bdc) is of great interest\cite{Pereiro2014}.

\section{Supporting Information}
The Supporting Information is available free of charge at https://pubs.acs.org.

Details about material preparation and characterization, ferromagnetic resonance measurement, SQUID measurement, data fittings, and theoretical calculations (PDF).

\medskip

\textbf{Corresponding emails:} jwen11@stanford.edu, yili@anl.gov, y.mokrousov@fz-juelich.de, weizhang@oakland.edu, lipeng18@ustc.edu.cn

\section{Acknowledgments}
\begin{acknowledgments}
Y.X. and W.Z. acknowledge the U.S. NSF under grant Grant Nos. ECCS-1933301 and ECCS-1941426. L.C., F.L, F.F. and Y.M. gratefully acknowledge the J\"ulich Supercomputing Centre and RWTH Aachen University for providing computational resources, as well as the support of Deutsche Forschungsgemeinschaft (DFG, German Research Foundation) - TRR 173 - 268565370 (project A11), TRR 288 – 422213477 (project B06). J.W. and Y.L. are supported by the U.S. Department of Energy (DOE), Office of Science, Basic Energy Sciences, Materials Sciences and Engineering Division, under contract DE-AC02-76SF00515. X.Z. acknowledges the support of the fellowship of China Postdoctoral Science Foundation No. 2021M701590. P.L. acknowledges the support of the U.S. NSF EPM Grant No. DMR-2129879, the Ralph E. Powe Junior Faculty Enhancement Award and discussions with Drs. Mingzhong Wu, Alexander Mook, Ran Cheng, and Hua Chen.

\end{acknowledgments}

%\section{Contributions}
%L.A., J.W. and P.L. conceived the work. J.W. and Y.S.L. prepared the samples. L.A., X.Z. M.M. and P.L. carried out the experimental measurements and data analysis. Y.X., Y.L., V.N., W.-K.K., D.Y., W.Z. contributed to the data analysis. L.Z., F.L., F.F., and Y.M. performed the theoretical analysis and calculations. P.L., Y.M., Y.S.L. and W.Z. supervised the project. L.A., L.Z., Y.M. and P.L. wrote the paper with the help from all authors.

\end{document}

\title{Supporting Information: Evidence of Magnon-Mediated Orbital Magnetism in a Quasi-2D Topological Magnon Insulator}

\author{Laith Alahmed}
\affiliation{Department of Electrical and Computer Engineering, Auburn University, Auburn, AL 36849, USA}

\author{Xiaoqian Zhang}
\affiliation{Shenzhen Institute for Quantum Science and Engineering, Southern University of Science and Technology, Shenzhen, 518055, China}

\author{Jiajia Wen}
\affiliation{Stanford Institute for Materials and Energy Sciences, SLAC National Accelerator Laboratory, Menlo Park, CA, 94025, USA}

\author{Yuzan Xiong}
\affiliation{Department of Physics, Oakland University, Rochester, MI 48309 USA} 

\author{Yi Li}
\affiliation{Materials Science Division, Argonne National Laboratory, Lemont, IL 60439}

\author{Li-chuan Zhang}
\affiliation{Peter Gr\"unberg Institut and Institute for Advanced Simulation,
Forschungszentrum J\"ulich and JARA, 52425 J\"ulich, Germany}

\author{Fabian Lux}
\affiliation{Institute of Physics, Johannes Gutenberg University Mainz, 55099 Mainz, Germany}

\author{Frank Freimuth} \affiliation{Peter Gr\"unberg Institut and Institute for Advanced Simulation,
Forschungszentrum J\"ulich and JARA, 52425 J\"ulich, Germany}
\affiliation{Institute of Physics, Johannes Gutenberg University Mainz, 55099 Mainz, Germany}

\author{Muntasir Mahdi}
\affiliation{Department of Electrical and Computer Engineering, Auburn University, Auburn, AL 36849, USA}

\author{Yuriy Mokrousov}
\affiliation{Peter Gr\"unberg Institut and Institute for Advanced Simulation,
Forschungszentrum J\"ulich and JARA, 52425 J\"ulich, Germany}
\affiliation{Institute of Physics, Johannes Gutenberg University Mainz, 55099 Mainz, Germany}

\author{Valentine Novosad}
\affiliation{Materials Science Division, Argonne National Laboratory, Lemont, IL 60439}

\author{Wai-Kwong Kwok}
\affiliation{Materials Science Division, Argonne National Laboratory, Lemont, IL 60439}

\author{Dapeng Yu}
\affiliation{Shenzhen Institute for Quantum Science and Engineering, Southern University of Science and Technology, Shenzhen, 518055, China}

\author{Wei Zhang}
\affiliation{Department of Physics, Oakland University, Rochester, MI 48309 USA}

\author{Young S. Lee}
\affiliation{Stanford Institute for Materials and Energy Sciences, SLAC National Accelerator Laboratory, Menlo Park, CA, 94025, USA}
\affiliation{Department of Applied Physics, Stanford University, Stanford, CA, 94305, USA}

\author{Peng Li}
\affiliation{Department of Electrical and Computer Engineering, Auburn University, Auburn, AL 36849, USA}

\maketitle

\section{Vibrating Sample Magnetometry} 
A Quantum Design DynaCool PPMS equipped with a VSM module was used to study the quasi-static magnetization properties of Cu(1,3-bdc). The single crystals have an average mass of $\sim$1 mg. The Dynacool system is capable of cooling samples down to 1.6 K and can generate a variable magnetic field up to $\pm$9 T. The VSM was used to measure the magnetic moment of Cu(1,3-bdc), in emu, while the magnetic field was swept at different temperatures. The magnetic field was applied along the OOP direction and along different IP angles 0$^\textrm{o}$, 90$^\textrm{o}$ and 120$^\textrm{o}$. The volume of the sample was calculated by using the mass and density of the sample. Magnetization was then extracted from the magnetic moment measurements by dividing by the sample volume. 
%(1 mT = 40$\pi\frac{emu}{cm^3}$).

\section{Broadband Ferromagnetic Resonance Spectroscopy} 
Broadband FMR spectroscopy was performed using the DynaCool PPMS CryoFMR module comprising a Co-Planar Waveguide (CPW), an external microwave source (BNC Model 845), and a lock-in amplifier (Signal Recovery Model 7265). The lock-in amplifier was used to record microwave power loss as an external field was swept IP and OOP of the sample, and at different microwave frequencies and temperatures. In addition, a vector network analyzer (FieldFox F9918A) was used to perform the FMR measurements and validate the results. The VNA both supplies a microwave signal and detects the returned S21 signal. The VNA measurements were repeated at different fixed IP and OOP field values and at different temperatures.

\section{Calculation of topological orbital moment}  
The TOM imprinted by magnons is calculated  via linear-spin-wave theory\cite{PhysRevB.87.144101,Malki_2020,Pereiro2014,Malki2019PRB,PhysRevB.100.144401} with the effective spin-wave Hamiltonian. The local TOM of the $n$th magnon branch in momentum
space is obtained according to $L_{n\mathbf k}^\mathrm{TOM}=\kappa^\mathrm{TO} \left\langle{\Psi_{n\mathbf k}|\chi(\mathbf k)|\Psi_{n\mathbf k}}\right\rangle$, where   $\kappa^{TO}$ is the topological orbital susceptibility, $|\Psi_{n\mathbf k}\rangle$ is the eigenstate of the $n$th magnon branch 
and $\chi(\mathbf k)$ is the scalar spin chirality in momentum space.
The summation of the TOM at given temperature $T$ is calculated with the formula: $\left\langle{L^\mathrm{TOM}} \right \rangle_T  = \sum_n \int\displaylimits_{\rm BZ} n_\mathrm{B}(\epsilon_{n\mathbf k}) \, L_{n\mathbf k}^\mathrm{TOM}\, d\mathbf k$, where the $n_\mathrm{B}(\epsilon_{n\mathbf k})$ represents the Bose distribution function with $n_\mathrm{B}(\epsilon_{n\mathbf k})=[\exp{(\epsilon_{n\mathbf k}/(k_\mathrm{B}T))}-1]^{-1}$ and $\epsilon_{{n\mathbf k}}$  is the eigenvalue of the $n$th magnon branch in the effective spin-wave Hamiltonian.  More details can be accessed in Ref.~\cite{zhang2020imprinting}. 

\section{In-plane VSM Measurements}
Fig. 1b in the main text has confirmed that Cu(1,3-bdc) favors an in-plane alignment slightly. The sample was rotated and measured at different IP angles: 0$^\textrm{o}$, 90$^\textrm{o}$, and 120$^\textrm{o}$ to examine if it has uniaxial anisotropy in-plane. Fig. S1 shows IP angular VSM measurements at 1.9 K, 5 K, and 9 K, respectively. At each temperature, the curves at different angles almost overlap, and no significant differences are observed. Therefore, we conclude that Cu(1,3-bdc) does not show uniaxial IP anisotropy.

\begin{figure}[h]
    \centering
    \includegraphics[width=1\columnwidth]{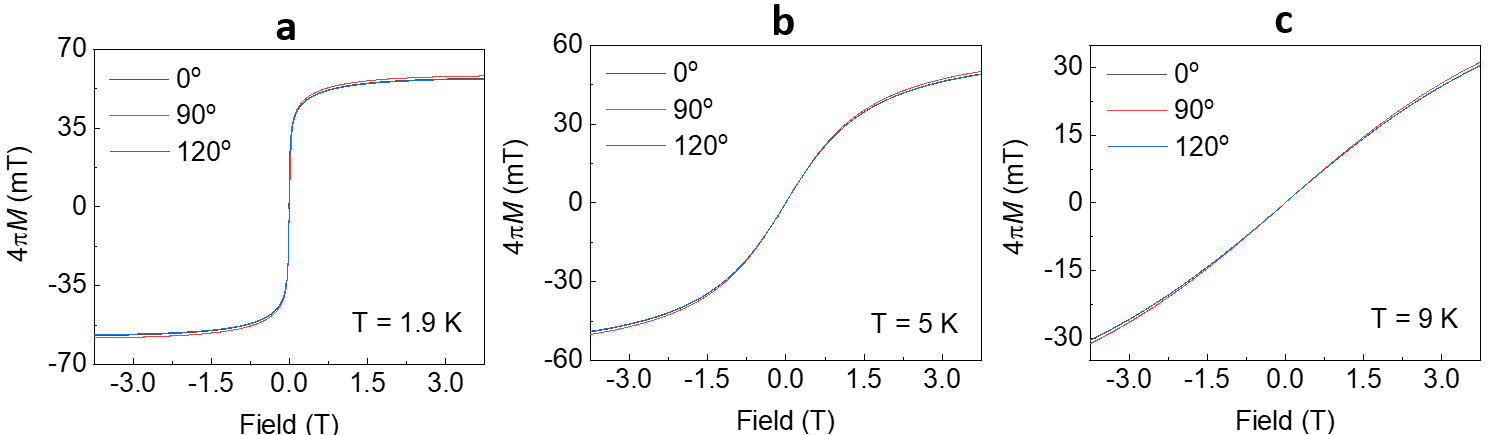}
    \caption{In-plane VSM measurements of Cu(1,3-bdc). \textbf{a.,b., and c.} VSM hysteresis loops at IP angles 0$^\textrm{o}$, 90$^\textrm{o}$, and 120$^\textrm{o}$ at 1.9 K, 5 K, and 9 K, respectively.}
    \label{figS1}
\end{figure}

\section{VSM data Fitting Procedure}
For an accurate extraction of the saturation magnetization $M_\textrm{s}$ values, the data collected from the VSM measurements were fitted to hyperbolic tangent functions as follows\cite{Li2021,PhysRevX.10.011012}:

\begin{equation}
M(H) = M_\textrm{s} tanh(\frac{H\pm H_\textrm{c}}{H_\textrm{0}}) \label{tanh}
\end{equation}

Where $H$ is the applied magnetic field, $M_\textrm{s}$ is the saturation magnetization, $H_\textrm{c}$ is the coercivity, and $H_\textrm{0}$ is a constant. Fig. S2 shows an example of such fitting; the VSM data collected at T = 1.9 K for an externally swept OOP field (red open circles) is fitted to Eq. (\ref{tanh}) (solid black line). All measured VSM data was fitted using Eq. (\ref{tanh}) to extract the $M_\textrm{s}$ values.

\begin{figure}[t]
    \centering
    \includegraphics[width=0.55\columnwidth]{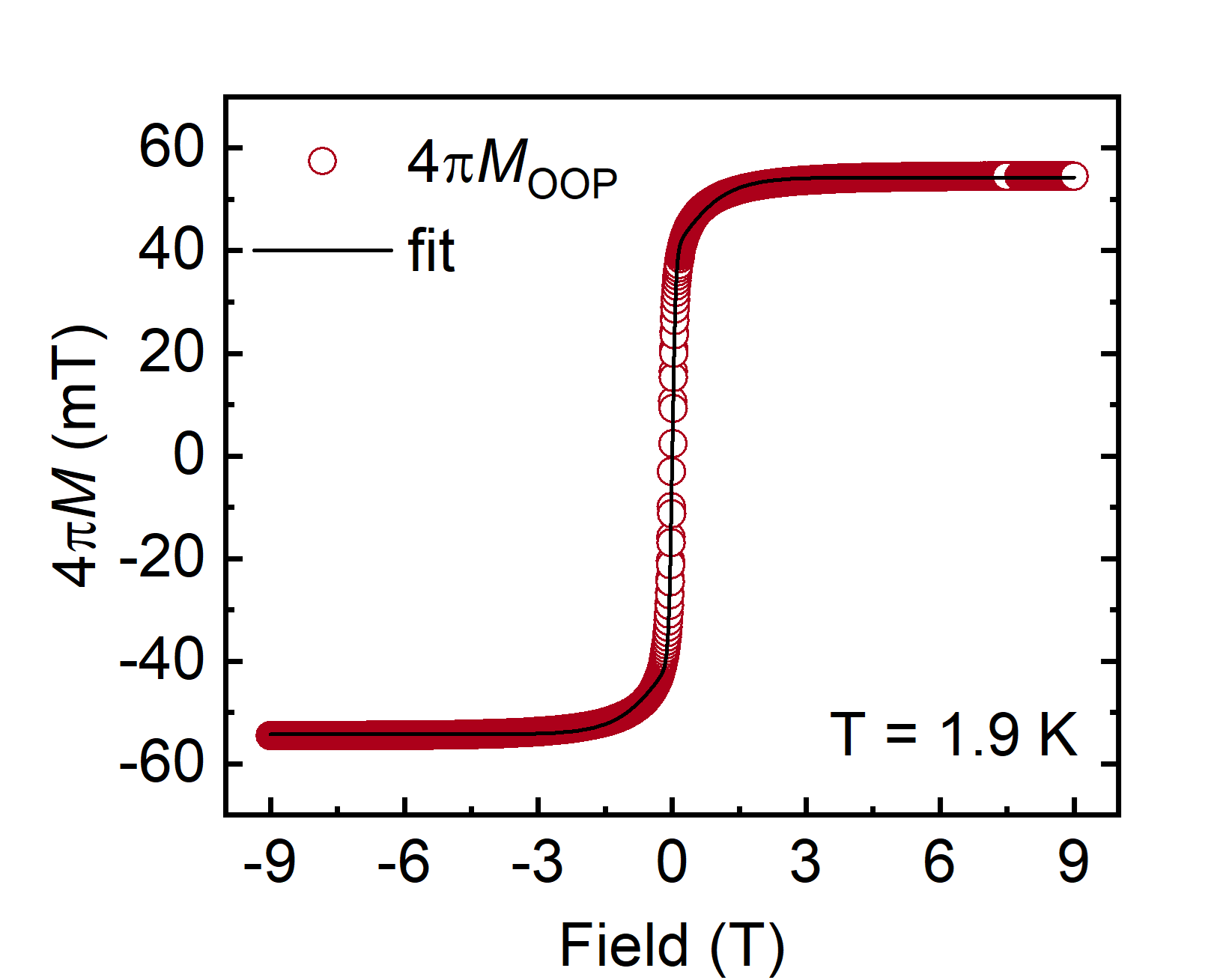}
    \caption{Fitted VSM measurement for an OOP field at T = 1.9 K. The circular data points represent the raw measured VSM data. The solid black line shows the fit to Eq. (1).}
    \label{figtanh}
\end{figure}

\section{FMR data Fitting Procedure}

The FMR profiles are fitted to the superposition of a Lorentzian + anti-Lorentzian functions\cite{Oates2002,LowYIG}. The fittings yield the resonance magnetic field and the Full Width at Half Maximum (FWHM) linewidth \(\Delta H_{\textrm{FWHM}}\). Then we plot the resonance frequencies \textit{f} vs. the FMR resonance fields \textit{H}$_\textrm{r}$ at different temperatures for the IP and OOP field directions in Fig. S3a. Next we follow the procedures described in Ref. \cite{LowYIG} to fit the resonance frequency vs. resonance field data points. Because Cu(1,3-bdc) is an easy plane magnet, the standard Kittel equations can be used to fit the data\cite{Oates2002,LowYIG}:

\begin{figure}[h]
    \centering
    \includegraphics[width=0.7\columnwidth]{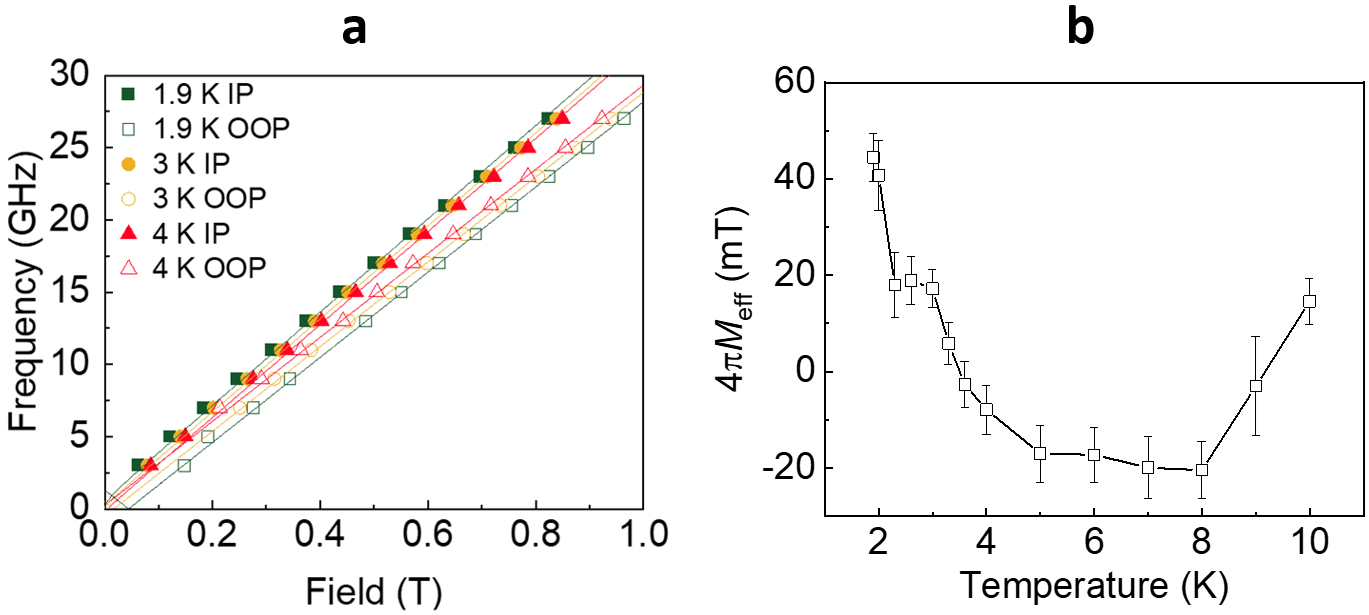}
    \caption{\textbf{a.} Resonance frequency \textit{f} vs. resonance field \textit{H}$_\textrm{r}$ at 1.9 K, 3 K, and 4 K for external fields applied IP (solid points) and OOP (open points) for Cu(1,3-bdc). The data points were fitted with Equations (2) and (3). \textbf{b.} Temperature dependence of FMR-extracted effective magnetization $4\pi M_\textrm{eff}$.}
    \label{figS2}
\end{figure}

\begin{equation}
f = \gamma'_\textrm{IP} \sqrt{(H_r +4\pi M_\textrm{eff}) H_r} \label{Kittel_IP}
\end{equation}
\begin{equation}
f = \gamma'_\textrm{OOP} (H_r -4\pi M_\textrm{eff}) \label{Kittel_OOP}
\end{equation}

where Equations (2) and (3) are standard equations for fitting IP and OOP data points, respectively. Here, $f$ is the microwave frequency, $\gamma'$ = $\frac{\gamma}{2\pi}$ is the reduced gyromagnetic ratio, $H_\textrm{r}$ is the resonance field, and $4\pi M_\textrm{eff}$ is the effective magnetization, which is dictated by the magnetocrystalline anisotropy field $H_\textrm{k}$. The fittings can yield $\gamma'$ and $4\pi M_\textrm{eff}$. As shown in Fig. \ref{figS2}b, $4\pi M_\textrm{eff}$ fluctuates between -20 mT and +50 mT within the tested temperature range (1.9 K to 10 K). To overcome the possible influence of $H_\textrm{k}$, one can analyze the data points measured at large magnetic fields to obtain accurate results. In this regard, Equations (2) and (3) can be re-written as the following:

\begin{equation}
f = \gamma'_\textrm{IP} (H_r +2\pi H_\textrm{eff})\label{Kittel_IP} \end{equation}
\begin{equation}
f = \gamma'_\textrm{OOP} (H_r -4\pi H_\textrm{eff}) \label{Kittel_OOP}
\end{equation}

where $H_\textrm{r}$ $>>$ $4\pi M_\textrm{eff}$. We replace $M_\textrm{eff}$ with $H_\textrm{eff}$ in the revised Equations. We have found that the standard equations (Eq. (2) and (3)) and the revised equations (Eq. (4) and (5)) generate the same results for our samples.

\section{Ferromagnetic Resonance with Vector Network Analyzer}

\begin{figure}[h]
    \centering
    \includegraphics[width=1\columnwidth]{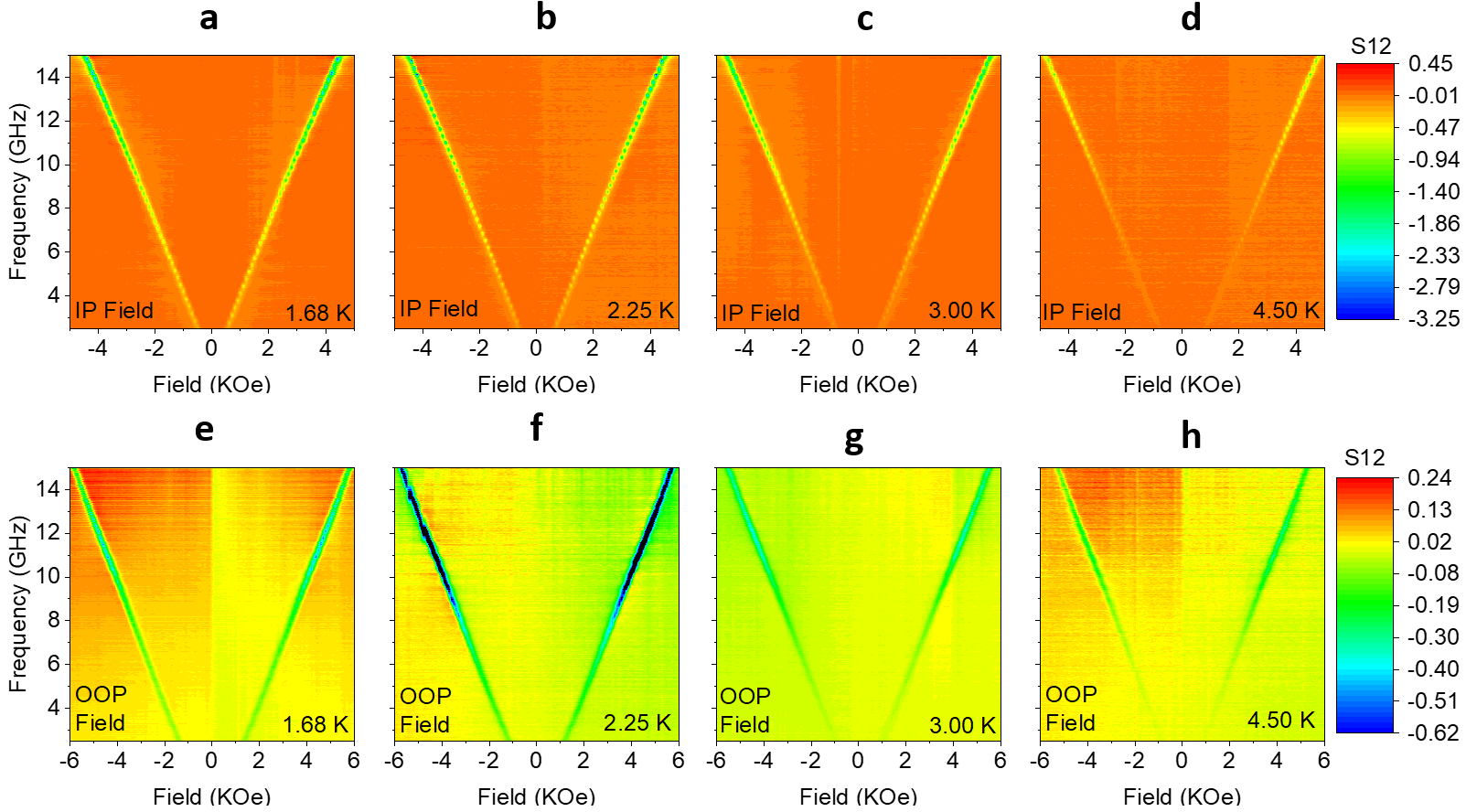}
    \caption{VNA FMR measurements of Cu(1,3-bdc) at different temperatures. \textbf{a.-d.} Frequency vs. IP field at 1.68 K, 2.25 K, 3.00 K, and 4.5 K, respectively. \textbf{e.-f.} Frequency vs. OOP field at 1.68 K, 2.25 K, 3.00 K, and 4.5 K, respectively. For all plots, the color represents the amplitude of the S12 signal.}
    \label{figS4}
\end{figure}

\begin{figure}[t]
    \centering
    \includegraphics[width=0.7\columnwidth]{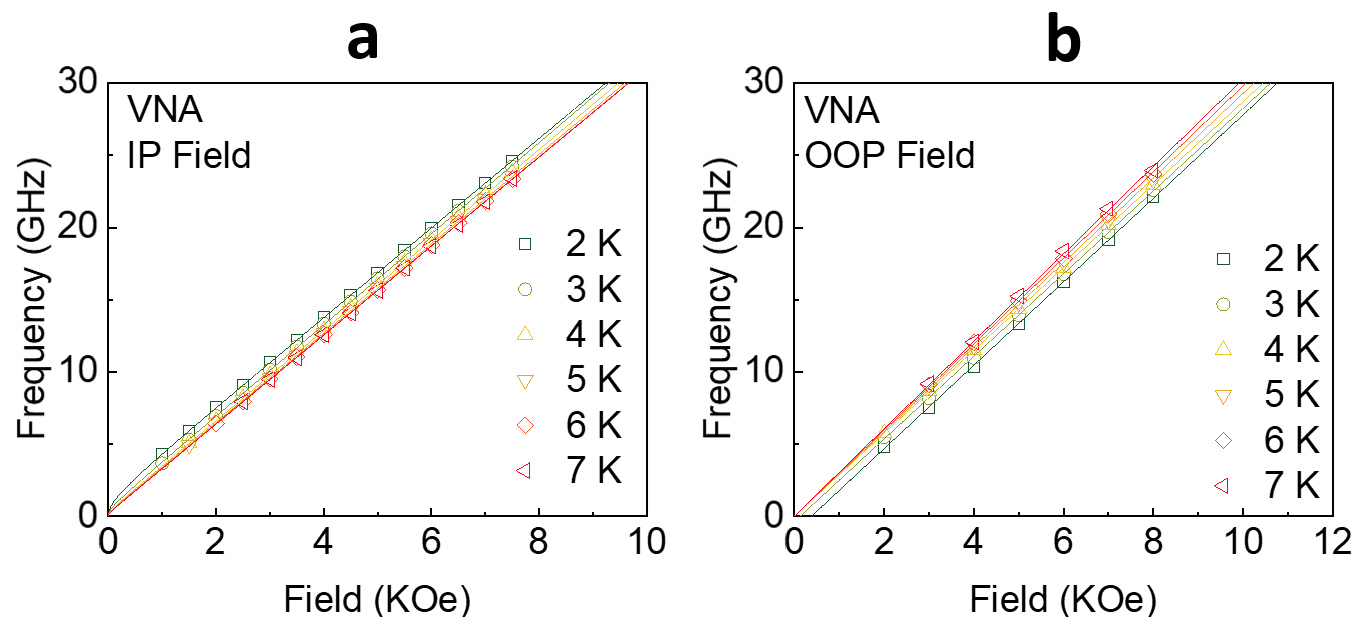}
    \caption{Kittel dispersion curves from the VNA measurements. \textbf{a.} The Frequency \textit{f} vs. resonance field \textit{H}$_\textrm{r}$ at 2 K, 3 K, 4 K, 5 K, 6 K, and 7 K for an externally applied IP field. The data points are fitted to Eq. (2). \textbf{b.} Frequency \textit{f} vs. resonance field \textit{H}$_\textrm{r}$ at 2 K, 3 K, 4 K, 5 K, 6 K, and 7 K for an externally applied OOP field. The data points are fitted to Eq. (3).}
    \label{figS5}
\end{figure}

\begin{figure}[h]
    \centering
    \includegraphics[width=0.8\columnwidth]{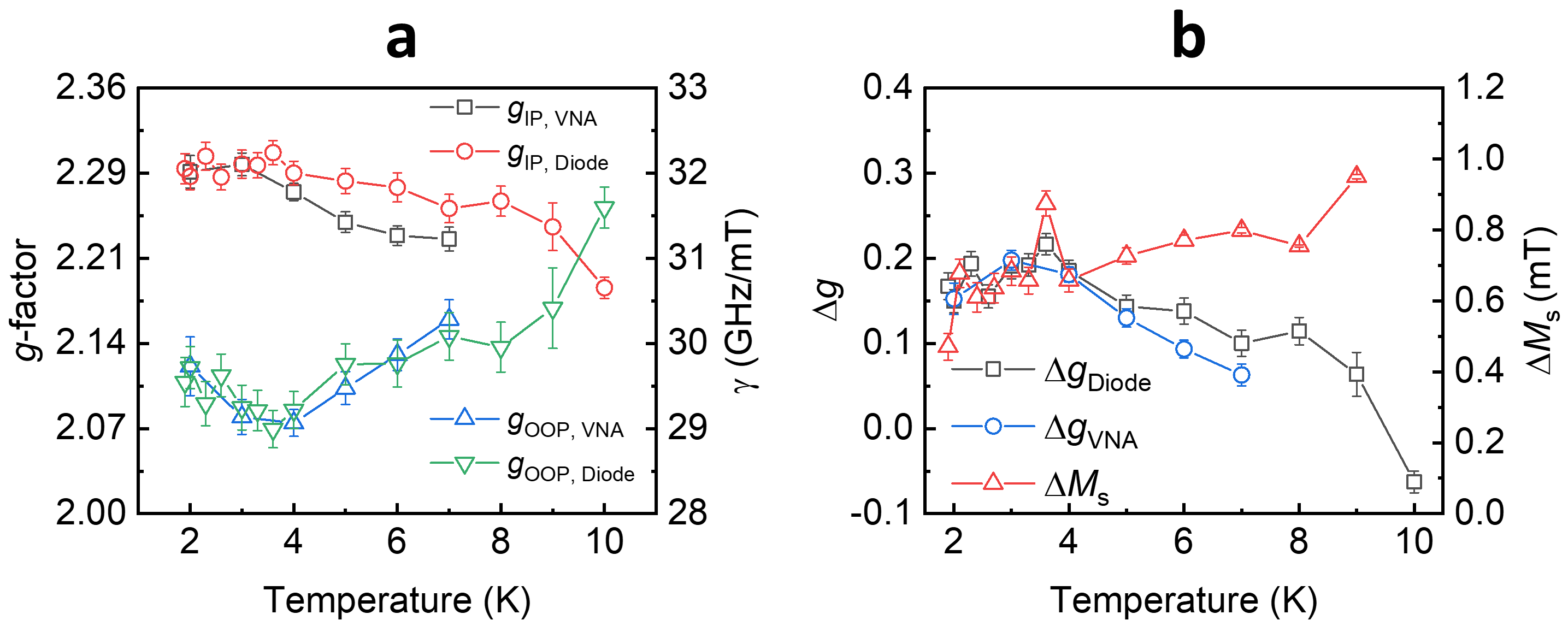}
    \caption{Comparison of VNA and microwave diode-based FMR data. \textbf{a.} Temperature dependence of the IP and OOP $g$-factor and gyromagnetic ratio $\gamma$ from both the VNA and microwave diode measurements. \textbf{b.} Correlation between VNA and microwave diode $\Delta$\textit{g} (plotted on the left axis), and $\Delta$ \textit{M}$_\textrm{s}$ (plotted on the right axis), as a function of temperature.}
    \label{figS6}
\end{figure}

In the main text, we show results of broadband Ferromagnetic Resonance (FMR) spectroscopy of Cu(1,3-bdc) performed using a microwave-diode based power absorption measurement technique. In such setup, the microwave frequency is fixed, and the external magnetic field is swept from high to low while the microwave absorption by the sample is measured. Another way to measure the sample is to use a Vector Network Analyzer (VNA). VNA directly measures the transmitted microwave signal S12 of the sample, where S12 refers to transmitted microwave signal from Port 1 to Port 2. In the VNA setup, the external magnetic field is fixed at a certain value and the frequency is scanned.

Figs S4 a-d (e-h) show the measured VNA response at different frequencies and external IP (OOP) magnetic field. In these plots, the color represents the S12 signal magnitude. Fig. S5 presents the extracted resonance frequency as a function of the resonance fields. The data points are fitted to Eq. (2) and (3) through the procedures illustrated in Note VI.

The gyromagnetic ratio and corresponding $g$-factor values, obtained from fitting the VNA data, show excellent agreement with the results acquired from the microwave diode-based measurement, as shown in Fig. S6a. In the figure, ``Diode" refers to the diode-based FMR measurements that were used in the main text, where a microwave diode was used to detect microwave absorption. As can be seen from Fig. S6a, the VNA and diode methods have yielded very similar $g$-factor and gyromagnetic ratio $\gamma$ values. Fig. S6b plots temperature dependence of $\Delta g$ and $\Delta M_\textrm{s}$. The results further show that the VNA and diode methods exhibit good agreement, where they both deviate from $\Delta M_\textrm{s}$ at higher temperatures.

\section{Analysis of FMR linewidth}

The FMR fittings discussed above as well as in the main text, also generate the Full-width-at-half-maximum (FWHM) linewidth $\Delta H_{\textrm{FWHM}}$. $\Delta H_{\textrm{FWHM}}$ is plotted as a function of temperature at different microwave frequencies, as shown in Fig. S7. $\Delta H_{\textrm{FWHM}}$ remains almost constant up to T $\sim$ 3 K, which is close to $T_\textrm{c}$ of Cu(1,3-bdc). As the temperature increases past 3 K, $\Delta H_{\textrm{FWHM}}$ starts to increase monotonically. The linewidth vs. temperature can be understood as follows: at a lower temperature (T \textless 4 K), Cu(1,3-bdc) becomes more ferromagnetic (long-range magnetic ordering), and the spin arrangement becomes more uniform. In addition, the magnon-phonon scattering at reduced temperature is lower. These two mechanisms lead to lower linewidths at lower temperatures. At higher temperatures (T \textgreater 4 K), the short-range magnetic ordering, together with increased magnon-phonon scattering, contribute to the broadened linewidths.

\begin{figure}[h]
    \centering
    \includegraphics[width=0.5\textwidth]{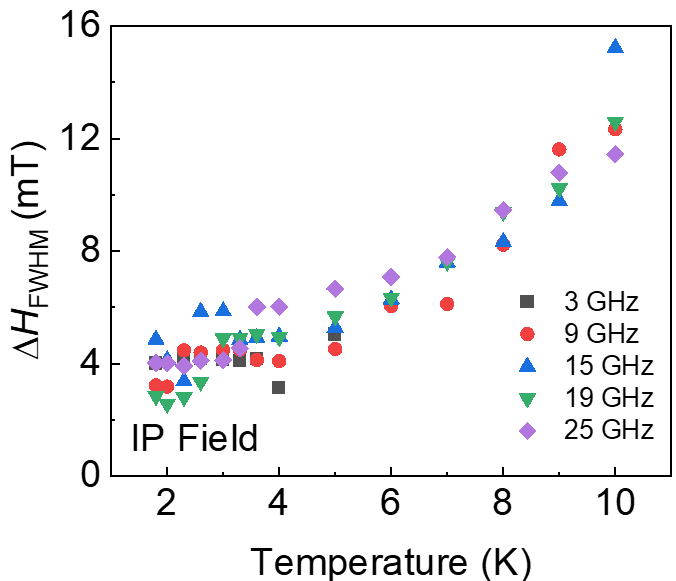}
    \caption{Full-width-at-half-maximum (FWHM) linewidth $\Delta H_\textrm{FWHM}$ vs. temperature at different microwave frequencies for IP field.}
    \label{figS7}
\end{figure}

\section{Sample growth and XRD Characterization}

The hydrothermal approach was used to grow the Cu(1,3-bdc) single crystals. More details regarding the sample preparation can be found in Ref.\cite{Nytko2008}. We have used x-ray diffraction (XRD) to understand the crystallinity of the sample. One can see that the (00l) peaks have been identified in the XRD spectrum in Fig. S8. We have estimated the d-spacing from the XRD peaks using d(hkl)=$\lambda$/(2sin$\theta$), which yields a d-spacing of 1.59 nm in the c-axis direction. This is consistent with the c-axis parameter of the Cu(1,3-bdc) unit cell reported in Ref. \cite{Nytko2008}.

\begin{figure}[h]
    \centering
    \includegraphics[width=0.5\textwidth]{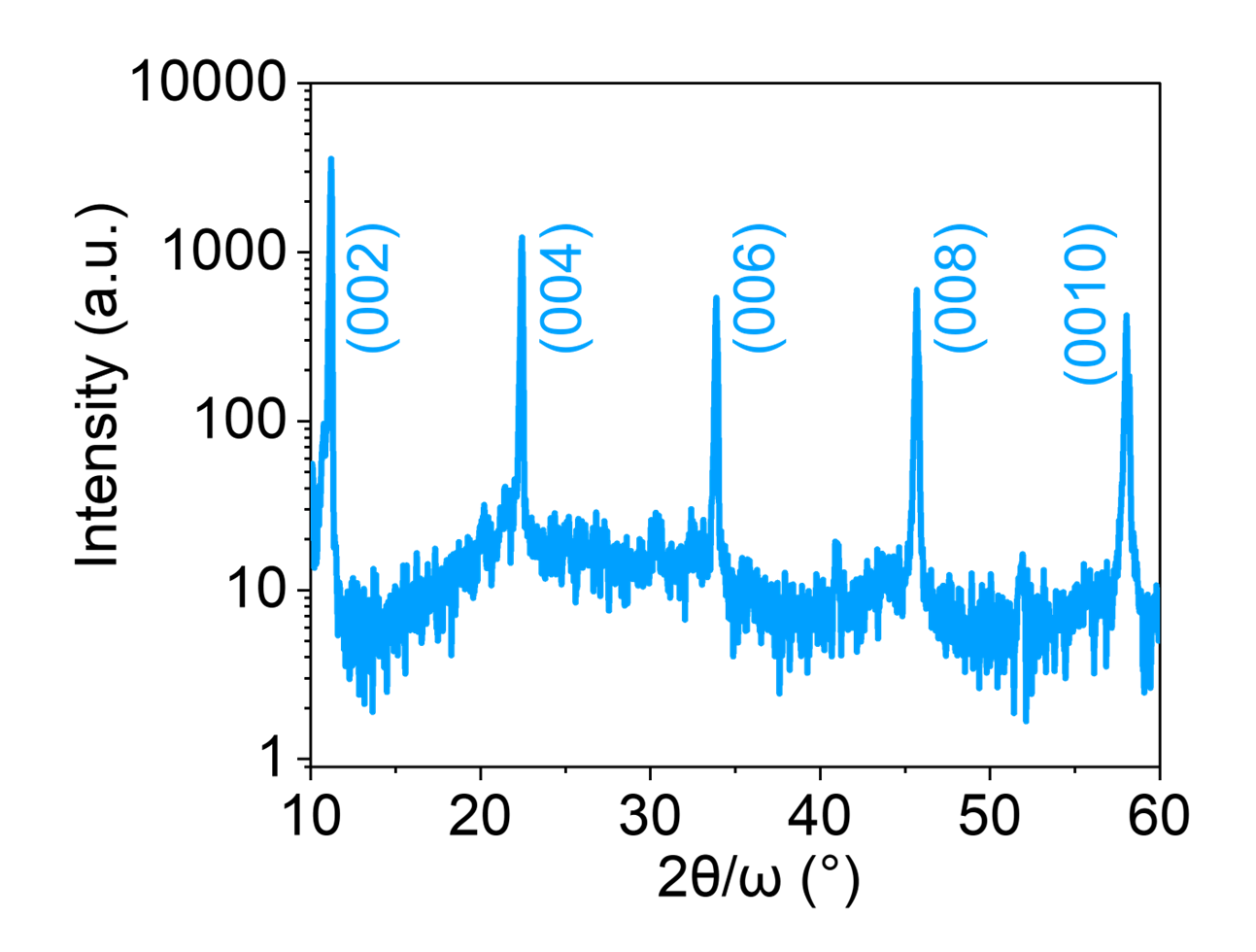}
    \caption{XRD spectrum of the Cu(1,3-bdc) crystal. The (00l) peaks indicates the c-axis orientation.}
    \label{figS7}
\end{figure}

\section{SQUID Measurement of magnetization}
We have carried out standalone IP and OOP M-H measurements in a Quantum Design Magnetic Property Measurement System (MPMS) SQUID system with higher sensitivity. Fig. S9a,b shows the IP and OOP M-H loops. Fig. S9c plots the temperature dependence of saturation magnetization vs. temperature in the IP and OOP directions measured by SQUID. Fig. S9d plots the new $\Delta$M$_\textrm{s}$(M$_\textrm{s,IP}$ - $M_\textrm{s,OOP}$) (SQUID data) and $M_\textrm{s,IP}$ - $M_\textrm{s,OOP}$ (VSM data). The temperature dependence of $\Delta$M$_\textrm{s}$ are consistent and show that $M_\textrm{s,IP}$ is higher than $M_\textrm{s,OOP}$ at all tested temperatures.

\begin{figure}[h!]
    \centering
    \includegraphics[width=0.8\textwidth]{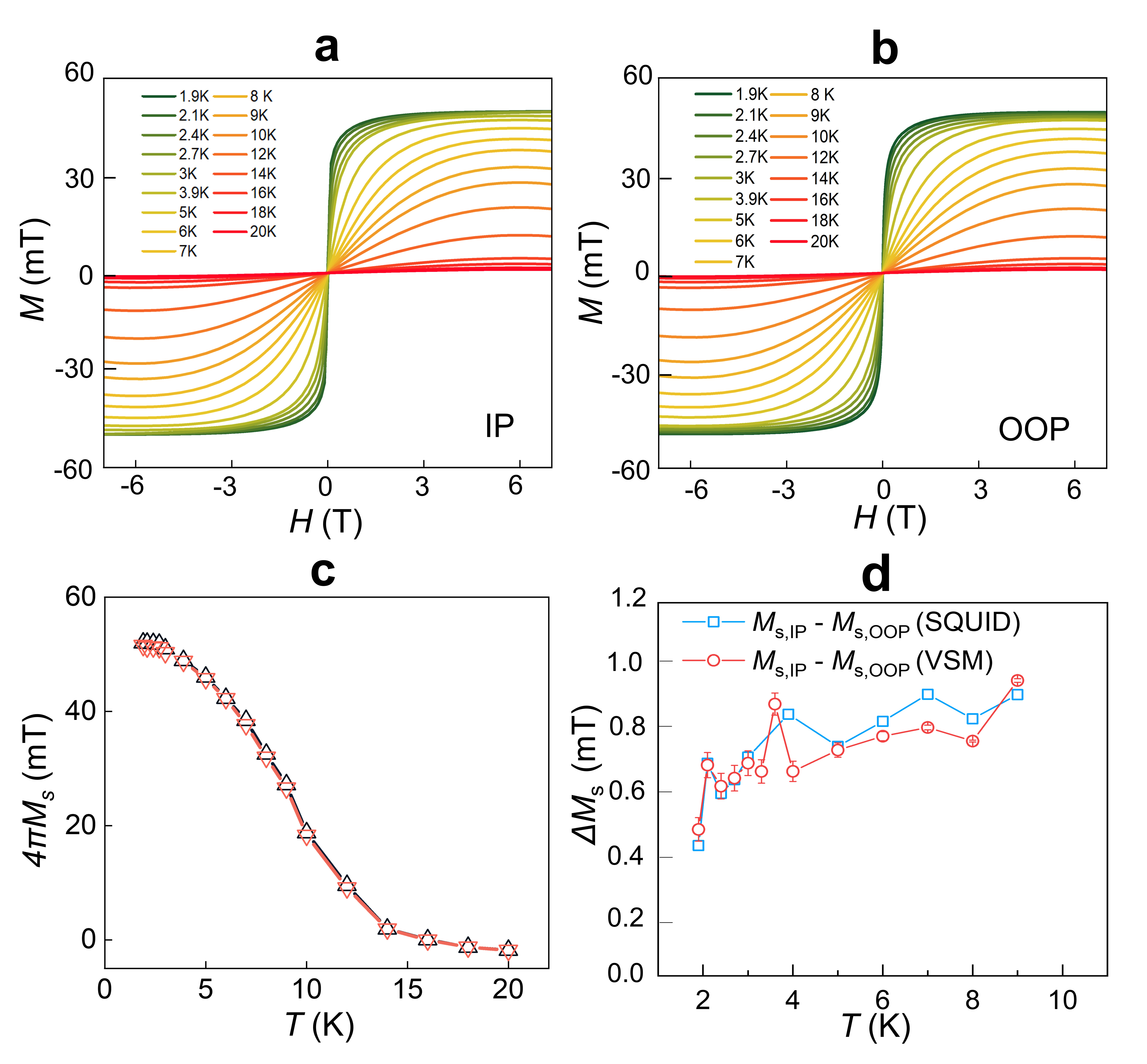}
    \caption{SQUID data and comparison with VSM data. a. SQUID IP $M-H$ loops. b. SQUID OOP $M-H$ loops. c. SQUID $M_\textrm{s}$ vs. $T$ curves. d. $\Delta$M$_\textrm{s}$($M_\textrm{s,IP}$ - $M_\textrm{s,OOP}$) vs. $T$ measured by both VSM and SQUID.}
    \label{figS7}
\end{figure}

\section{Temperature dependent FMR measurement}

We have plotted the FMR profiles at 21 GHz at temperatures ranging from 2 K to 10 K in the figure below. It should be noted that the magnitude of the FMR signals degrade with increasing temperatures, which is a result of reduced magnetization at higher temperatures. 

\begin{figure}[h!]
    \centering
    \includegraphics[width=0.8\textwidth]{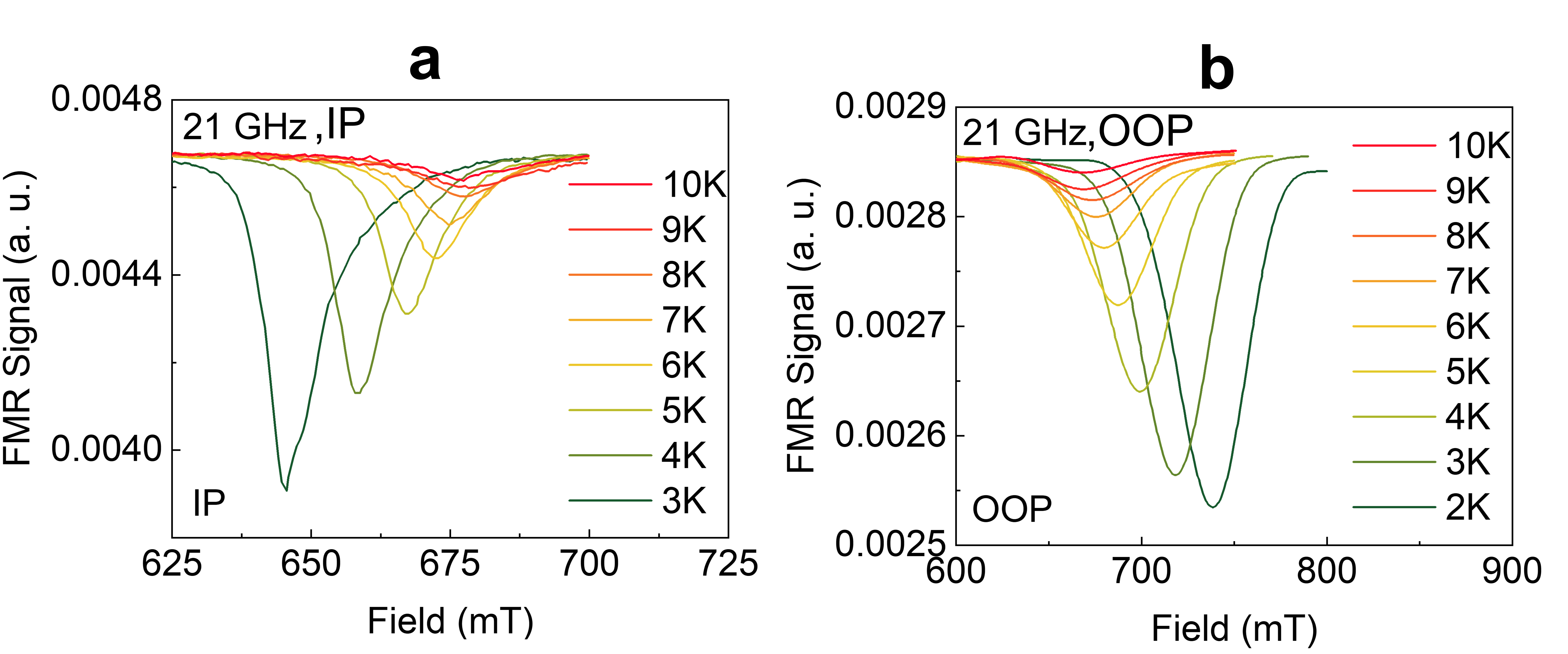}
    \caption{Ferromagnetic resonance profiles at 21 GHz. a. In-plane curves. b. Out-of-plane curves.}
    \label{figS7}
\end{figure}

\maketitle

\centering